\documentclass[aps, prb, twocolumn, showpacs] {revtex4}
\usepackage{epsfig,amsfonts,amssymb,amsmath}

\begin{document}

%%%%%%%%%%%%%%%%%%%%%%%%%%%%%%%%%%%%%%%%%%%%%%%%%%%%%%%%%%%%%%%%%
% Defintions   
%%%%%%%%%%%%%%%%%%%%%%%%%%%%%%%%%%%%%%%%%%%%%%%%%%%%%%%%%%%%%%%%%
\renewcommand{\Re}{\operatorname{Re}}
\renewcommand{\Im}{\operatorname{Im}}
\newcommand{\Tr}{\operatorname{Tr}}
\newcommand{\sign}{\operatorname{sign}}
\newcommand{\dd}{\text{d}}
\newcommand{\q}{\boldsymbol q}
\newcommand{\p}{\boldsymbol p}
\newcommand{\rr}{\boldsymbol r}
\newcommand{\pp}{p_v}
\newcommand{\vv}{\boldsymbol v}
\newcommand{\I}{{\rm i}}
\newcommand{\pphi}{\boldsymbol \phi}
\newcommand{\ds}{\displaystyle}
\newcommand{\be}{\begin{equation}}
\newcommand{\ee}{\end{equation}}
\newcommand{\bea}{\begin{eqnarray}}
\newcommand{\eea}{\end{eqnarray}}
\newcommand{\Acl}{{\cal A}}
\newcommand{\Rcl}{{\cal R}}
\newcommand{\Tcl}{{\cal T}}
\newcommand{\Tmin}{{T_{\rm min}}}
\newcommand{\Toff}{{\langle \delta T \rangle_{\rm off} }}
\newcommand{\Roff}{{\langle \delta R \rangle_{\rm off} }}
\newcommand{\RoffI}{{\langle \delta R_I \rangle_{\rm off} }}
\newcommand{\RoffII}{{\langle \delta R_{II} \rangle_{\rm off} }}
\newcommand{\dg}{{\langle \delta g \rangle_{\rm off} }}
\newcommand{\rd}{{\rm d}}
\newcommand{\br}{{\bf r}}
\newcommand{\la}{\langle}
\newcommand{\ra}{\rangle}
\newcommand{\ua}{\uparrow}
\newcommand{\da}{\downarrow}
%%%%%%%%%%%%%%%%%%%%%%%%%%%%%%%%%%%%%%%%%%%%%%%%%%%%%%%%%%%%%%%%%%%

\title{Geometric Phases and Andreev Reflection in Hybrid Rings} 

\author{Diego Frustaglia, Fabio Taddei, and Rosario Fazio}  
\affiliation{
NEST-INFM \& Scuola Normale Superiore, 56126 Pisa, Italy}

\date{\today}

\begin{abstract}
We study the Andreev reflection of a hybrid mesoscopic ring in the presence of a crown-like 
magnetic texture. By calculating the linear-response conductance as a function of the 
Zeeman splitting and the magnetic flux through the ring, we are able to identify 
signatures of the Berry phase acquired by the electrons during transport. This is 
proposed as a novel detection scheme of the spin-related Berry phase, having the advantage 
of a larger signal contrast and robustness against ensemble averaging.
\end{abstract} 


% 03.65.Vf Phases: geometric; dynamic or topological
% 72.10.-d Theory of electronic transport; scattering mechanisms
% 72.10.Bg General formulation of transport theory
% 73.21.-b Electron states and collective excitations in multilayers, 
%          quantum wells, mesoscopic, and nanoscale systems 
% 73.23.-b Electronic transport in mesoscopic systems
% 74.45.+c Proximity effects; Andreev effect; SN and SNS junctions
% 85.75.-d Magnetoelectronics; spintronics: devices exploiting spin 
%          polarized transport or integrated magnetic fields

\maketitle

%%%%%%%%%%%%%%%%%%%%%%%%%%%%%%%%%%%%%%%%%%%%%%%%%%%%%%%%%%%%%%%%%%%%%
% BODY OF PAPER
%%%%%%%%%%%%%%%%%%%%%%%%%%%%%%%%%%%%%%%%%%%%%%%%%%%%%%%%%%%%%%%%%%%%%

\section{Introduction}

Ferromagnet-superconductor (FS) hybrid systems are an attractive subject 
of research because of the competition between the spin asymmetry characteristic 
of a ferromagnet (due to the spin-splitting induced by the exchange field) and the correlations 
(occurring among electrons belonging to opposite spin species) induced by superconductivity. 
At low energies, electronic transport in mesoscopic FS hybrid systems is dominated by Andreev 
reflection \cite{A64}. For excitation energies below the superconducting gap $\Delta$, 
single electrons coming from the ferromagnet cannot penetrate into the superconductor. 
Nevertheless, a current can flow due to the Andreev reflection process: A spin-up particle 
in the majority band is reflected at the FS interface back into the minority band as a 
spin-down hole, leading to the formation of a Cooper pair in the superconductor. 
Andreev reflection, however, is suppressed by increasing the exchange field of the 
ferromagnet up to the limiting case of a half-metal, where only one spin species has a 
finite density of states and the current vanishes. Starting from the earlier experimental \cite{fierz90,petrashov94} 
and theoretical \cite{JB95} investigations, the research activity in this subject has increased 
rapidly in the last few years \cite{FSreviews05} due to the interest in spintronics \cite{ZFS04}.

If the ferromagnet contains a tunnel insulating (I) barrier, to form a FIFS structure, 
multiple scattering takes place giving rise to 
quasi-bound states that can be resolved in the {\it linear-response} conductance by 
resonant Andreev tunneling (RAT) \cite{JB95}. This is a quantum interference effect 
between particles and holes of opposite spin owning different wave vectors because of 
the relative Zeeman splitting. Another important quantum interference effect in hybrid 
structures is the Andreev interference, which occurs in systems with at least two 
superconducting regions kept at different superconducting-order-parameter 
phases \cite{Andreevinter}.

For {\it nonuniform} magnetizations (magnetic {\it textures}) further interference is
expected due to the presence of additional spin-related phases of geometric origin, usually 
referred to as Berry phases \cite{B84}. 
Berry phases arise when the carriers spin adiabatically follow the magnetic texture and 
stay (anti)align with it during transport. They manifest as an effective spin-dependent magnetic 
flux of geometric origin (geometric flux) and their magnitude is proportional to the corresponding 
solid angle accumulated by the spins. In mesoscopic physics, Berry phases have been extensively 
investigated by considering their effects in normal Aharonov-Bohm (AB) interferometers 
\cite{LGB90,S92,ALG93,FR01,FHR01} and multiple efforts have 
been done for detection \cite{YTW98,JG98,MHKWB98,YPS02,R99,MMKKN04,YYLG04}. 

%%%%%%%%%%%%%%%%%%%%%%%%%%%%%%%%%%%%%%%%%%%%%%%%%%%%%%%%%%%%%%%%%%%%%%%%%%%%%%%%%%%%%%%%%%
%                                       FIGURE
%%%%%%%%%%%%%%%%%%%%%%%%%%%%%%%%%%%%%%%%%%%%%%%%%%%%%%%%%%%%%%%%%%%%%%%%%%%%%%%%%%%%%%%%%%
\begin{figure}
%[tbp]
\includegraphics[width=.4 \textwidth, angle=0]{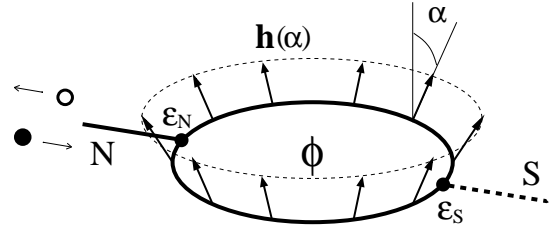}
\caption{Hybrid magnetic-superconducting ring setup. 
The ring is subject to a magnetic texture ${\bf h}(\alpha)$ 
which couples to the carriers spin. 
The coupling points of the ring (denoted by small filled circles) are
attached to a normal lead (on the left-hand-side) and to a 
superconducting lead (on the right-hand-side). Andreev reflection is the only contribution to the subgap conductance $G_{\rm MS}$. 
}
\label{fig-1}
\end{figure}
%%%%%%%%%%%%%%%%%%%%%%%%%%%%%%%%%%%%%%%%%%%%%%%%%%%%%%%%%%%%%%%%%%%%%%%%%%%%%%%%%%%%%%%%%%

In this paper we bring together these two distinct physical phenomena (Andreeev reflection 
and Berry phases) by studying the Andreev conductance of an hybrid one-dimensional (1D) 
mesoscopic ring in the presence of a magnetic texture (Fig.~\ref{fig-1}). The proposed setup 
permits to identify the signatures of magnetic and geometric phases in the energy spectrum 
of hybrid ring geometries \cite{note8}.
Our work can be considered an alternative proposal for the detection of Berry phases: 
Andreev reflection has the advantage of including particle-hole phase 
correlations that allow for a larger signal contrast and a high 
sensitivity to the magnetic/geometric flux-dependent splitting of quasibound states.
Unlikely the case of normal systems, this is true even for ensemble-averaged quantities.

\section{Andreev reflection and magnetic textures} 

Let us introduce the Bogoliubov-de Gennes (BdG) equation for particle (p) and 
hole (h) spinors ($\Psi_{\rm p}$ and $\Psi_{\rm h}$, respectively).
This reads \cite{deGennes}
\be
\left( \begin{array}{cc}
H_0\,\b{1}+{\bf h} \cdot \underline{\boldsymbol{\sigma}} & \underline{\Delta}\\
-\underline{\Delta}^* & -(H_0\,\b{1}+{\bf h} \cdot \underline{\boldsymbol{\sigma}})^*
\end{array} \right)
\left( \begin{array}{c}
\Psi_{\rm p}\\
\Psi_{\rm h}
\end{array} \right)
=E
\left( \begin{array}{c}
\Psi_{\rm p}\\
\Psi_{\rm h}
\end{array} \right),
\label{BdG}
\ee
where $E$ is the quasiparticle energy measured from the condensate chemical potential 
$\mu=E_{\rm F}$ equal to the Fermi energy, 
$H_0=\boldsymbol{\Pi}^2/2m+V-\mu$ is the single-particle Hamiltonian with generalized 
momentum $\boldsymbol{\Pi}={\bf p}+(e/c){\bf A}_{\rm em}$ ($e > 0$) and impurity potential 
$V({\bf r})$, ${\bf h}({\bf r})$ is a position-dependent exchange field describing a 
magnetic texture, $\underline{\boldsymbol{\sigma}}$ is the Pauli matrix vector,
\be
\underline{\Delta}({\bf r})=
\left( \begin{array}{cc}
0 & \Delta({\bf r}) \\
-\Delta({\bf r}) & 0
\end{array} \right),
\ee
where $\Delta({\bf r})$ is the particle-hole coupling field (superconducting order parameter), and $\b{1}$ is the (2$\times$2) unit matrix.
The BdG equation (\ref{BdG}) can be used for studying transport in magnetic-superconducting 
(MS) hybrid junctions, where FS junctions with uniform magnetization are just a particular 
case. Locating the MS interface at $x = 0$, we assume a stepwise superconducting order parameter
$\Delta({\bf r}) = \Theta(x) \Delta_0$ \cite{note1}. 
The conductance of the hybrid system can be calculated by using a scattering approach.
For energies $0 \le E < \Delta_0$ the only contribution to the conductance is 
given by Andreev reflection. 
However, while in usual FS junctions an incoming spin-up particle is eventually reflected 
as a spin-down hole, in MS systems the situation is different: Due to the presence of spin-flip, an incoming spin-up particle can be reflected as a hole either with spin up or down, 
opening an additional channel for Andreev reflection \cite{note10}.
The zero-temperature scattering formula for the subgap linear-response conductance $G_{\rm MS}$ 
of a single-channel MS hybrid structure is given by \cite{BTK82,L91,TE92,JB95} 
\be
G_{\rm MS}=\frac{2 e^2}{h} \sum_{\sigma'\sigma} |r_{\rm hp}^{\sigma'\sigma}|^2,
\label{Gms}
\ee
where $r_{\rm hp}^{\sigma'\sigma}$ is the Andreev reflection amplitude for incoming particles with spin 
$\sigma$ to outgoing holes with spin $\sigma'$ calculated at $E=0$. 
The Andreev reflection matrix $(\b{r}_{\rm hp})_{\sigma'\sigma}\equiv r_{\rm hp}^{\sigma'\sigma}$ can be written in terms of the scattering amplitudes of the system when the superconducting lead is replaced by a normal lead \cite{JB95,B92}:
\be
\b{r}_{\rm hp}=\b{t}'_{\rm h}\b{$\varrho$}_{\rm hp}[\b{1}+\b{r}'_{\rm p}\b{$\varrho$}_{\rm ph}\b{r}'_{\rm h}\b{$\varrho$}_{\rm hp}]^{-1}\b{t}_{\rm p},
\label{ra1}
\ee
where, at the Fermi energy,
\be
\b{$\varrho$}_{\rm hp}=
\left( \begin{array}{cc}
0 & i \\
-i & 0
\end{array} \right)
\ee
is the Andreev-reflection amplitude matrix at a perfect NS interface for incoming particles  
($\b{$\varrho$}_{\rm ph}=\b{$\varrho$}_{\rm hp}^*$ for incoming holes). 
In Eq.~(\ref{ra1}), $\b{t}_{\rm p}$ and $\b{t}'_{\rm h}$ refer, respectively, to particle left-to-right and hole right-to-left transmission amplitudes, while $\b{r}'_{\rm p(h)}$ refers to particle (hole) right-to-right reflection amplitude.
Eq. (\ref{ra1}) can be further simplified by using the particle-hole symmetry of the S-matrix 
and writing the hole amplitudes in terms of particle ones. 

\section{Andreev reflection in a hybrid MS ring}

We now apply the above formulation for the study of
the hybrid ring geometry shown in Fig.~\ref{fig-1}. The setup consists of a ballistic single-channel metallic ring of radius $a$ subject to a crown-like magnetic texture 
${\bf h}(\alpha)$ characterized by the tilt angle $\alpha$.
The field ${\bf h}(\alpha)$ can either be generated by a micromagnet 
\cite{YTW98,JG98} or describe the exchange field in a ferromagnetic 
ring \cite{RKLVBBCS01,HDKVB03}. The magnetic 
texture owns a solid angle $\Omega_{\rm g}(\alpha)=2 \pi(1-\cos \alpha)$ which give rise to 
geometric phases in the regime of adiabatic spin transport \cite{B84,note2}. In addition, 
a magnetic flux 
$\phi$ is also allowed through the ring. For the sake of clarity ${\bf h}(\alpha)$ and $\phi$ will 
be considered independent. This permits to separate the effect of magnetic phases
originated by $\phi$, from geometric phases depending on $\alpha$.     
Moreover, the ring is coupled to a normal lead (N) at the left and a superconducting lead (S) at 
the right.
The coupling to each lead is described by a three-terminal scattering-matrix model 
\cite{BIA84,HSFR04} where incoming quasiparticles from the left-hand-side (right-hand-side) lead 
are transmitted into the two branches of the ring with equal probability 
$0 \le \varepsilon_{\rm N}(\varepsilon_{\rm S}) \le 1/2$. Reflection occurs with probabilities 
$1-2\varepsilon_{\rm N}$ and 
$1-2\varepsilon_{\rm S}$. In particular, for $\varepsilon_{\rm N}=\varepsilon_{\rm S}=0$, 
the ring is completely isolated from the leads.
We calculate the Andreev reflection amplitude (\ref{ra1}) by using a spin-dependent transfer-matrix 
approach \cite{HSFR04} generalized to the case of asymmetric coupling to the leads 
($\varepsilon_{\rm N} \neq \varepsilon_{\rm S}$).

By adjusting the coupling to the leads it is possible to go from the 
RAT regime (for $\varepsilon_{\rm N} \ll \varepsilon_{\rm S} \approx 1/2$)
to the resonant normal tunneling (RNT) regime (for $\varepsilon_{\rm N}, \varepsilon_{\rm S} \ll 1/2$).
On the one hand, RAT reflects the presence of Andreev quasibound states determined by the multiple particle-hole scattering, occurring between the contacts, and represented by the factor $[\b{1}+\b{r}'_{\rm p}\b{$\varrho$}_{\rm ph}\b{r}'_{\rm h}\b{$\varrho$}_{\rm hp}]^{-1}$ in Eq.~(\ref{ra1}).
On the other hand, RNT is controlled by the ordinary quasibound states determined by the ring itself (being weakly coupled to both leads) and represented by the transmission amplitudes $\b{t}_{\rm p}$ and $\b{t}'_{\rm h}$. 
In both RAT and RNT regimes quasibound states are resolved in the linear-response conductance $G_{\rm MS}$.

We study $G_{\rm MS}$ for different field and geometry 
configurations by calculating the normal transmission and reflection amplitudes present 
in Eq.~(\ref{ra1}). 
By setting $\varepsilon_{\rm N} = 0.1$ ($\ll 1/2$) we force 
Andreev-reflected particles and holes to be trapped within the ring for some time before escaping, 
favoring the development of quasibound states. It is convenient to characterize the effective Zeeman splitting $|{\bf h}|$ by introducing
\be
Q= a \delta k, 
\label{Q}
\ee
with $\delta k= k^\da-k^\ua$, $k^{\da,\ua}=k_{\rm F}\sqrt{1 \pm z}$, $z=|{\bf h}|/E_{\rm F}$, and $E_{\rm F}=\hbar^2/2m k_{\rm F}^2$.
The quantity $2\pi Q$ is the relative phase accumulated between up particles and down holes in a single round trip around the ring because of their different wave vectors.  
In the limit $z \ll 1$ (where most interesting physics happens \cite{note7}) we can write 
$Q \approx \ell z$, with $\ell= k_{\rm F} a$ the dimensionless angular momentum of the carriers. 
Additional spin-related phases arise exclusively from the nonuniformity of the 
magnetic texture ($\alpha \neq 0$).
In the limit in which the spins adiabatically follow the magnetic texture during transport 
the corresponding wave functions accumulate a geometric Berry phase. This depends only on $\alpha$ and the spin orientation $\sigma=\ua,\da$ {\it relative to the local field}. 
The Berry phase manifests itself in the form of an 
effective geometric vector potential ${\bf A}^\sigma_{\rm g}(\alpha)$ 
to be added to ${\bf A}_{\rm em}$ in the kinetic term 
of Eq.~(\ref{BdG}). This forces the carriers to experience an {\it extra} spin-dependent flux 
$\phi^\sigma_{\rm g}(\alpha)=\sigma \phi_{\rm g}(\alpha)$, where 
$\phi_{\rm g}(\alpha)\equiv \phi_0 \Omega_{\rm g}(\alpha)/4\pi=\phi_0(1- \cos \alpha)/2$ and 
$\phi_0 \equiv hc/e$ (flux quantum).
Therefore each spin-carrier feels a total effective flux 
$\phi+\sigma \phi_{\rm g}$ \cite{LGB90,S92,FR01}.  
The amount of geometric flux through the ring is characterized by 
$\phi_{\rm g}$, which goes from zero, for $\alpha = 0$, to $\phi_0$, for $\alpha = 180^\circ$. 
For the setup of Fig.~\ref{fig-1} the adiabatic limit corresponds to the condition:
$Q_\alpha \equiv Q/|\sin \alpha| \gg 1$ \cite{note4}, so that adiabatic spin transport can be guaranteed for all $\alpha$ by setting $Q \gg 1$ \cite{note9}.
Taking into account that $G_{\rm MS}$ is periodic in $Q$ (at least 
for ballistic systems with $z \ll 1$ \cite{JB95}), we can formally limit our discussion to 
small values of $Q$ without loss of generality \cite{note5}.

%%%%%%%%%%%%%%%%%%%%%%%%%%%%%%%%%%%%%%%%%%%%%%%%%%%%%%%%%%%%%%%%%%%%%%%%%%%%%%%%%%%%%%%%%%
%                                       FIGURE
%%%%%%%%%%%%%%%%%%%%%%%%%%%%%%%%%%%%%%%%%%%%%%%%%%%%%%%%%%%%%%%%%%%%%%%%%%%%%%%%%%%%%%%%%%
\begin{figure}
%[tbp]
\includegraphics[width=.45 \textwidth, angle=0]{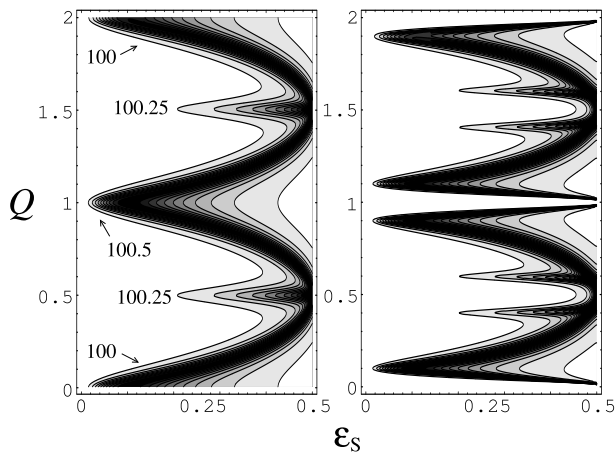}
\caption{$G_{\rm MS}$ vs $Q$ and $\varepsilon_{\rm S}$ for a weakly coupled N lead 
($\varepsilon_{\rm N}=0.1$), $\phi=0$, and three different values of energy 
corresponding to $\ell=$ 100, 100.25 and 100.5. The left panel  
shows results for a uniform texture ($\alpha=0$, $\phi_{\rm g}=0$). In the right panel 
a nonuniform texture ($\alpha \approx 26^\circ$) introduces a small geometric flux 
$\phi_{\rm g}=0.05 \phi_0$ leading to a splitting of the pattern.
}
\label{fig-2}
\end{figure}
%%%%%%%%%%%%%%%%%%%%%%%%%%%%%%%%%%%%%%%%%%%%%%%%%%%%%%%%%%%%%%%%%%%%%%%%%%%%%%%%%%%%%%%%%%

\section{Results and discussion}

\begin{figure}
%[tbp]
\includegraphics[width=.35 \textwidth, angle=0]{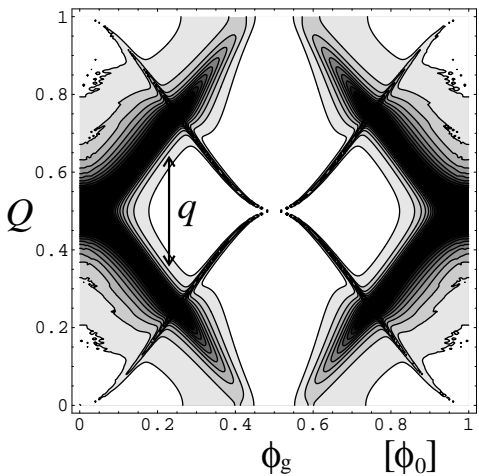}
\caption{$\la G_{\rm MS}\ra_\ell$ vs $Q,\phi_{\rm g}$ for $\phi=0$. 
The coupling $\varepsilon_{\rm N}=0.1$ and $\varepsilon_{\rm S}=1/2$ favors RAT. 
Resonances show a splitting $q$ a function of $\phi_{\rm g}$.
}
\label{fig-3}
\end{figure}

\begin{figure}
%[tbp]
\includegraphics[width=.35 \textwidth, angle=0]{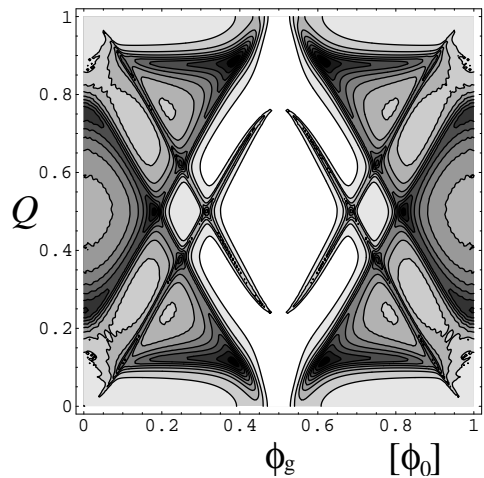}
\caption{Idem Fig.~\ref{fig-3} now for a partial coupling $\varepsilon_{\rm S}=0.4$.
}
\label{fig-4}
\end{figure}

\begin{figure}
%[tbp]
\includegraphics[width=.35 \textwidth, angle=0]{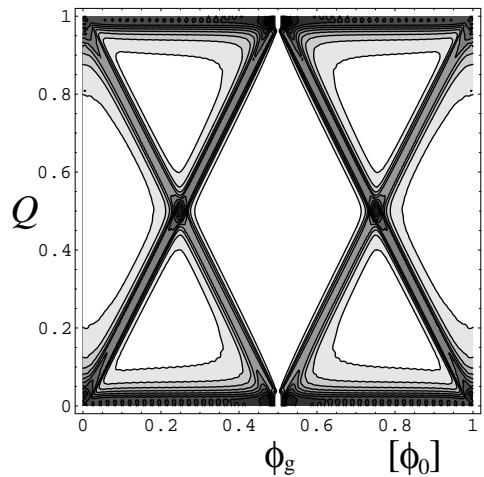}
\caption{Idem Fig.~\ref{fig-3} now with small $\varepsilon_{\rm S}=\varepsilon_{\rm N}=0.1$ 
(weakly coupled ring). This favors RNT.
}
\label{fig-5}
\end{figure}

\begin{figure}
%[tbp]
\includegraphics[width=.35 \textwidth, angle=0]{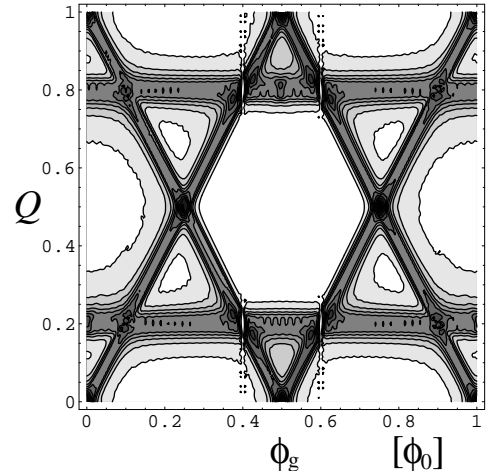}
\caption{$\la G_{\rm MS}\ra_\ell$ vs $Q,\phi_{\rm g}$ for a finite 
$\phi=0.1 \phi_0$, to be compared with Fig.~\ref{fig-5}. 
The additional splitting allows for a separation of the $\phi$ and $\phi_{\rm g}$
contributions.
}
\label{fig-6}
\end{figure}

From now on we present the results of our calculations for the Andreev conductance of hybrid rings 
and discuss the signatures of spin-related phases. The results are shown in a series of density 
plots (Figs.~\ref{fig-2}-\ref{fig-6}) for the normalized $G_{\rm MS}$, running from 0 (white) 
to 1 (black).   

We first discuss how different tunneling regimes arise by varying the coupling 
$\varepsilon_{\rm S}$ 
and consider the simplest case of a uniform texture ($\alpha=0$), where geometric 
phases are absent. 
The normal conductance of a partially coupled ballistic ring depends strongly on Fermi 
energy due to the presence of quasibound states, oscillating between maxima for integer 
$\ell$ (resonance with an eigenstate of the closed ring) and minima for half-integer $\ell$ 
(out of resonance) \cite{BIA84}. 
Analogous features are observed in the conductance of 
a hybrid ballistic ring, as illustrated 
in Fig.~\ref{fig-2}(left) where we plot $G_{\rm MS}$ for three different 
values of $\ell=$~100,~100.25~and~100.5 as a function of $Q$ and $\varepsilon_{\rm S}$
with $\phi=0$.
For perfect coupling ($\varepsilon_{\rm S}=1/2$) narrow {\it Andreev quasibound states} 
develop, producing large RAT peaks in the conductance $G_{\rm MS}$ at half-integer values of $Q$, corresponding to a relative accumulated-phase difference between up particles and down holes of $\pi$, independently of $\ell$.
These resonances are dominated by the second order contribution to $\b{r}_{\rm hp}$, in Eq.~(\ref{ra1}), corresponding to the round trip which includes three Andreev reflections at the S contact and two normal reflections at the N contact. 
Moreover, the fact that $G_{\rm MS}$ is independent of $\ell$ reflects the absence of ordinary bound state in the ring.
These behaviors are analogous to the case of a planar FIFS junction of width $L=\pi a$.\cite{JB95}

For $\varepsilon_{\rm S} < 1/2$, the peaks relative to the different values of $\ell$ separate as the carriers start to suffer normal reflection at the right-hand-side coupling point of the ring \cite{note3}. As $\varepsilon_{\rm S}$ is diminished further, the S lead gradually decouples from the ring, eventually reaching the RNT regime resulting from the formation of {\it ordinary quasibound states}. 
For $\varepsilon_{\rm S}\sim 0.1$, $G_{\rm MS}$ shows peaks only for integer and half-integer values of $\ell$, now at integer $Q$.
These behaviors can be understood as follows.
Finite contributions to $G_{\rm MS}$ comes from a sequence of two coherent processes. 
The first one consists in a spin-up particle tunneling through the ring from left to right via a resonant level (with amplitude $t_{\rm p}^\ua$). The second process consists of a spin-down Andreev reflected holes at the S interface tunneling through the ring from right to left via a resonant level (with amplitude ${t'}_{\rm h}^\da$).
The Andreev reflection amplitude $\b{r}_{\rm hp}$ is finite only if, at the Fermi energy, the two levels coincide (spin-degeneracy condition).
In the absence of any flux this happens only at integer values of $Q$. 

One can expect the conductance features of Fig.~\ref{fig-2}(left) to be sensitive to additional quantum phases introduced in the system.
In closed hybrid rings, for example, Andreev and ordinary bound state energies are found to split for a finite flux $\phi$ \cite{BK86,CKM03}.
In Fig.~\ref{fig-2}(right), which corresponds to Fig.~\ref{fig-2}(left) in the presence of a small geometric flux $\phi_{\rm g}=0.05 \phi_0$ (nonuniform texture with $\alpha \approx 26^\circ$), it is shown that all the resonances are indeed split.

In order to get a more general picture, independent of the particular value of the 
Fermi energy, it is convenient to average $G_{\rm MS}$ with respect to $\ell$ over a range $\delta \ell \gtrsim 1$.
We shall denote such an average by $\la G_{\rm MS}\ra_\ell$.
This corresponds either to an average on the kinetic energy of the incoming particles in an 
energy window of the order of the mean level spacing of the ring or, alternatively, to a sample
average for an ensemble of rings of different size. Note that it is {\it not} equivalent, 
as in the case of normal systems, to a finite temperature since the average is performed on the reference energy, $E_{\rm F}=\mu$, and not on the excitation energy $E$ which is fixed to zero. 
In a normal system, such an average makes any trace of resonant transport in the conductance unlikely, since $\delta \ell$ is larger than the necessary resolution. 
For a hybrid system as considered in this work, the situation is different: For $\ell \gg \delta \ell$ it is $\la Q \ra_\ell \approx Q$ 
and corrections $\delta Q \approx Q \delta \ell/\ell$, leading to dephasing between particles
and holes with opposite spin, can be neglected. 
This is very interesting since it means that, in this limit, phase correlations leading to narrow 
Andreev reflection resonances are kept. This outstanding 
property shown by hybrid setups appears as a great advantage over normal ones.
It permits the 
study of spectral features that otherwise could remain inaccessible. 

We note, in Fig.~\ref{fig-2}(left), that while for a given $\ell$ the conductance $G_{\rm MS}$ has a periodicity 2 as a function of $Q$, the plot including three values of $\ell$ shows periodicity 1 in $Q$. We shall therefore expect a period halving in the energy-averaged $G_{\rm MS}$.

%%%%%%%%%%%%%%%%%%%%%%%%%%%%%%%%%%%%%%%%%%%%%%%%%%%%%%%%%%%%%%%%%%%%%%%%%%%%%%%%%%%%%%%%%%
%                                       FIGURE
%%%%%%%%%%%%%%%%%%%%%%%%%%%%%%%%%%%%%%%%%%%%%%%%%%%%%%%%%%%%%%%%%%%%%%%%%%%%%%%%%%%%%%%%%%
%%%%%%%%%%%%%%%%%%%%%%%%%%%%%%%%%%%%%%%%%%%%%%%%%%%%%%%%%%%%%%%%%%%%%%%%%%%%%%%%%%%%%%%%%%

%%%%%%%%%%%%%%%%%%%%%%%%%%%%%%%%%%%%%%%%%%%%%%%%%%%%%%%%%%%%%%%%%%%%%%%%%%%%%%%%%%%%%%%%%%
%                                       FIGURE
%%%%%%%%%%%%%%%%%%%%%%%%%%%%%%%%%%%%%%%%%%%%%%%%%%%%%%%%%%%%%%%%%%%%%%%%%%%%%%%%%%%%%%%%%%
%%%%%%%%%%%%%%%%%%%%%%%%%%%%%%%%%%%%%%%%%%%%%%%%%%%%%%%%%%%%%%%%%%%%%%%%%%%%%%%%%%%%%%%%%%

%%%%%%%%%%%%%%%%%%%%%%%%%%%%%%%%%%%%%%%%%%%%%%%%%%%%%%%%%%%%%%%%%%%%%%%%%%%%%%%%%%%%%%%%%%
%                                       FIGURE
%%%%%%%%%%%%%%%%%%%%%%%%%%%%%%%%%%%%%%%%%%%%%%%%%%%%%%%%%%%%%%%%%%%%%%%%%%%%%%%%%%%%%%%%%%
%%%%%%%%%%%%%%%%%%%%%%%%%%%%%%%%%%%%%%%%%%%%%%%%%%%%%%%%%%%%%%%%%%%%%%%%%%%%%%%%%%%%%%%%%%

Results for $\la G_{\rm MS}\ra_\ell$, averaged over $100 \le \ell \le 102$, as a function of $Q$ and $\phi_{\rm g}$ in the absence of magnetic flux ($\phi = 0$) are organized in 
Figs.~\ref{fig-3}, \ref{fig-4}, and \ref{fig-5} for different coupling $\varepsilon_{\rm S}$. 
For $\varepsilon_{\rm S}= 1/2$, Fig.~\ref{fig-3}, 
we find a splitting $q=2 \phi_{\rm g}/\phi_0$ of the original RAT peaks located at 
half-integer $Q$ (see Fig.~\ref{fig-2}). In addition, "new" 
narrow RAT peaks with similar splitting and relative shift $\phi_0/2$ show up for finite 
$\phi_{\rm g}$. They have origin in the integer and half-integer $\ell$-branches 
of Fig.~\ref{fig-2}(left) which suffer an abrupt splitting for a finite flux, 
Fig.~\ref{fig-2}(right).
For a partial (more realistic) coupling $\varepsilon_{\rm S}= 0.4$, Fig.~\ref{fig-4}, 
we first observe that the RAT peaks split as a function of $\phi_{\rm g}$ at different 
points due to the $\ell$-dependent splitting already present for zero flux, Fig.~\ref{fig-2}(left). 
More interesting is, however, the 
increase of the slope $|\partial q/\partial (\phi_{\rm g}/\phi_0)| > 2$. 
This tendency reaches its maximum for small $\varepsilon_{\rm S}= 0.1$, Fig.~\ref{fig-5}, where the 
ring is weakly coupled to both leads and RNT dominates. There we see well defined branches which 
split with a slope $|\partial q/\partial (\phi/\phi_0)| = 4$, doubling that of 
Fig.~\ref{fig-3}. We note that Fig.~\ref{fig-5} reproduce the spin-degeneracy points of 
a closed normal ring.

It is interesting to notice that plots equal to those presented in Figs.~\ref{fig-3}, \ref{fig-4} and \ref{fig-5} can be obtained for a uniform magnetic texture ($\alpha=0$), in the presence of a finite magnetic flux $\phi$, by replacing $\phi_{\rm g} \rightarrow \phi$ in the horizontal axis.
However, for a nonuniform magnetic texture an effective interplay between the two phases $\phi$ and $\phi_{\rm g}$ takes place.
This case can be implemented by 
using the complete spin-dependent effective flux $\phi+\sigma \phi_{\rm g}$ 
in the normal amplitudes 
of Eq.~(\ref{ra1}), leading to a rich pattern which does {\it not} correspond to a simple 
shift along the flux axis. 
Instead, a multiple splitting arise in both $Q$ and flux axis, which
actually permits a clear distinction between magnetic and geometric flux contributions.
As an example we show in Fig.~\ref{fig-6} results for a weakly coupled ring  
($\varepsilon_{\rm S} = \varepsilon_{\rm N} = 0.1$) where we plot 
$\la G_{\rm MS}\ra_\ell$ as a function of $Q$ and $\phi_{\rm g}$ in the presence of a finite 
magnetic flux $\phi = 0.1 \phi_0$, to be compared with Fig.~\ref{fig-5}. 
The distinctive features are the following. The vertical line of vanishing $G_{\rm MS}$ present in Fig.~\ref{fig-5} for $\phi_{\rm g}=\phi_0/2$, in the presence of a magnetic flux is split into two lines occurring at $\phi_{\rm g}=\phi_0/2 \pm \phi$ (see Fig.~\ref{fig-6}). 
This happens because the transmission amplitudes $t_{\text{p}}^{\ua}$ and 
$t_{\text{h}}^{\prime\da}$, in Eq.~(\ref{ra1}), vanish at different values of $\phi_{\rm g}$, 
namely at $\phi_{\rm g}=\phi_0/2-\phi$ for spin-up particles and $\phi_{\rm g}=\phi_0/2+\phi$ for 
spin-down holes.
Moreover, we find that the horizontal resonance lines occurring at integer $Q$ in Fig.~\ref{fig-5}, 
in the presence of a magnetic flux appear at integer values of $Q\pm 2 \phi/\phi_0$ 
(see Fig.~\ref{fig-6}). This reflects the fact that the spin-degeneracy condition depends on both 
geometric and magnetic fluxes. We further notice that the plot in Fig.~\ref{fig-6} is invariant 
under the interchange $\phi_{\rm g} \leftrightarrow \phi$. Namely, calculating 
$\la G_{\rm MS}\ra_\ell$ vs $Q,\phi$ for a finite $\phi_{\rm g}=0.1 \phi_0$ 
($\alpha \approx 37^\circ$) give rise to an 
equal plot where $G_{\rm MS}$ now vanishes at vertical lines $\phi=\phi_0/2 \pm \phi_{\rm g}$ 
and horizontal resonance lines appear at integer values of $Q\pm 2 \phi_{\rm g}/\phi_0$. Such 
depiction is more appropiate for possible experimental realizations where the magnetic 
texture is weakly dependent on the external field producing the magnetic flux $\phi$. 

Finally, we note that in general the Zeeman splitting $Q$ will be a function of the applied 
flux $\phi$. However, the behavior of the Andreev conductance as a function of $\phi$ for a
varying $Q$ can still be seen from Figs.~\ref{fig-3}-\ref{fig-6}. One simply needs to specify 
a relationship between the applied $\phi$ and $Q$ (which would depend on $\alpha$), defining a 
path in the ($Q$,$\phi$)-plane. Moreover, $Q$ is not expected to vary much 
in the scale of one quantum of flux. Hence, for a given total flux, small 
variations around it should keep $Q$ constant.

%%%%%%%%%%%%%%%%%%%%%%%%%%%%%%%%%%%%%%%%%%%%%%%%%%%%%%%%%%%%%%%%%%%%%%%%%%%%%%%%%%%%%%%%%%
%                                       FIGURE
%%%%%%%%%%%%%%%%%%%%%%%%%%%%%%%%%%%%%%%%%%%%%%%%%%%%%%%%%%%%%%%%%%%%%%%%%%%%%%%%%%%%%%%%%%
%%%%%%%%%%%%%%%%%%%%%%%%%%%%%%%%%%%%%%%%%%%%%%%%%%%%%%%%%%%%%%%%%%%%%%%%%%%%%%%%%%%%%%%%%%

\section{Comments and Conclusion} 

We studied signatures of spin-related geometric and magnetic phases in the linear-response 
Andreev conductance $G_{\rm MS}$ of hybrid mesoscopic rings subject to a magnetic texture. 
$G_{\rm MS}$ appears to be very sensitive to both the single and combined effect of 
geometric and magnetic phases, even after an ensemble average.
This constitutes a promising alternative proposal for the detection of 
spin-related geometrical phases and their interplay with AB magnetic ones. 
We focused on ballistic systems. However, the introduction of weak 
disorder does not change significantly the results, leading mainly to some loss of 
contrast in the $G_{\rm MS}$ density plots \cite{FTF}. Similar effects are observed when
considering asymmetric ballistic rings where the arms present a (small) difference in 
length \cite{FTF}.  
Moreover, the conductance features found can be smoothed away by some phase averaging arising 
for large $Q$, specially in the presence of disorder \cite{JB95}. The latter can also risk the assumption of adiabatic spin transport \cite{PFR03}. 
However, non-adiabatic Aharonov-Anandan phases \cite{AA87} can still lead to similar results 
\cite{FTF}. 

We finally mention that recently developed mesoscopic magnetic structures based 
on paramagnetic n-doped diluted-magnetic-semiconductors in combination with micromagnets \cite{dietl} and ferromagnetic rings \cite{RKLVBBCS01,HDKVB03} could be consider as possible candidate systems.

\acknowledgments 

We thank G. Tkachov for a helpful clarification. This work has been supported by the 
EU Spintronics Research Training Network.

%%%%%%%%%%%%%%%%%%%%%%%%%%%%%%%%%%%%%%%%%%%%%%%%%%%%%%%%%%%%%%%%%%%%%
% BIBLIOGRAPHY
%%%%%%%%%%%%%%%%%%%%%%%%%%%%%%%%%%%%%%%%%%%%%%%%%%%%%%%%%%%%%%%%%%%%%

%%%%%%%%%%%%%%%%%%%%%%%%%%%%%%%%%%%%%%%%%%%%%%%%%%%%%%%%%%%%%%%%%%%%%


\begin{thebibliography}{99}
\bibitem{note8} Note the difference between the hybrid geometry of Fig.~\ref{fig-1} and those 
studied previously \cite{BK86,LL95,CKM03}, where a section of a closed normal ring is replaced 
by a superconductor. We instead attach a normal ring to a superconducting lead.
\bibitem{note7} Note that for $z \gtrsim 1$ one of the spin components is filtered (half-metallic 
ferromagnet) and the Andreev reflection contribution to the conductance vanishes \cite{JB95}.
\bibitem{note9} We note that adiabatic spin transport does not necessarily implies $Q \gg 1$, 
since for small $\alpha$ the actual condition $Q_\alpha \gg 1$ can be satisfied also for a 
small $Q$.
\bibitem{note5} We further note that in a ballistic ring with $\alpha=90^\circ$ the adiabatic limit 
is well satisfied already for $Q \approx 5$ \cite{FHR01}.
\bibitem{note3} Actually, for $\varepsilon_{\rm S}=1/2$ there is already a finite probability for a carrier to scatter back into the ring without suffering Andreev reflection at the S lead. However, in this particular case the scattering is highly symmetric maximizing the probability of Andreev reflection.
\end{thebibliography}
\end{document}